\documentclass[aps,pra,twocolumn,superscriptaddress,amsmath,amssymb]{revtex4}
\usepackage{graphicx,graphics}
\usepackage{amsmath,sidecap}
\usepackage{color}
\begin{document}
\title{Topologically nontrivial and trivial flat bands via weak and strong interlayer coupling in twisted bilayer honeycomb optical lattices for ultracold atoms}
\author{Wenjie Sui}
\affiliation{State Key Laboratory of Quantum Optics Technologies and Devices, \\  Institute of Opto-Electronics, Collaborative Innovation Center of Extreme Optics, Shanxi University, Taiyuan, Shanxi 030006, China}
\author{Wei Han}
\email[]{hanwei.irain@gmail.com}
\affiliation{State Key Laboratory of Quantum Optics Technologies and Devices, \\  Institute of Opto-Electronics, Collaborative Innovation Center of Extreme Optics, Shanxi University, Taiyuan, Shanxi 030006, China}
\affiliation{Liaoning Academy of Materials, Shenyang 110167, China}
\author{Zheng Vitto Han}
\affiliation{State Key Laboratory of Quantum Optics Technologies and Devices, \\  Institute of Opto-Electronics, Collaborative Innovation Center of Extreme Optics, Shanxi University, Taiyuan, Shanxi 030006, China}
\affiliation{Liaoning Academy of Materials, Shenyang 110167, China}
\author{Zengming Meng}
\affiliation{State Key Laboratory of Quantum Optics Technologies and Devices, \\  Institute of Opto-Electronics, Collaborative Innovation Center of Extreme Optics, Shanxi University, Taiyuan, Shanxi 030006, China}
\affiliation{Hefei National Laboratory, Hefei 230088, China}
\author{Jing Zhang}
\email[]{jzhang74@sxu.edu.cn}
\affiliation{State Key Laboratory of Quantum Optics Technologies and Devices, \\  Institute of Opto-Electronics, Collaborative Innovation Center of Extreme Optics, Shanxi University, Taiyuan, Shanxi 030006, China}
\affiliation{Hefei National Laboratory, Hefei 230088, China}
\date{\today}

\begin{abstract}
In recent years, flat electronic bands in twisted bilayer graphene (TBG) have attracted significant attention due to their intriguing topological properties, extremely slow electron velocities, and enhanced density of states. Extending twisted bilayer systems to new configurations is highly desirable, as it offers promising opportunities to explore flat bands beyond TBG. Here, we study both topological and trivial flat bands in a twisted bilayer honeycomb lattice for ultracold atoms and present the evolution of the flat bands with different interlayer coupling strength (ICS). Our results demonstrate that an isolated topological flat band can emerge at the Dirac point energy for a specific value of weak ICS, referred to as the ``critical coupling". This occurs over a wide range of twist angles, surpassing the limits of the magic angle in TBG systems. When the ICS is slightly increased beyond the critical coupling value, the topological flat band exhibits degenerate band crossings with both the upper and lower adjacent bands at the high-symmetry $\Gamma_s$ point. As the ICS is further increased into the strong coupling regime, trivial flat bands arise around Dirac point energy. Meanwhile, more trivial flat bands appear, extending from the lowest to higher energy bands, and remain flat as the ICS increases. The topological properties of the flat bands are studied through the winding pattern of the Wilson loop spectrum. Our research provides deeper insights into the formation of flat bands in ultracold atoms with highly controllable twisted bilayer optical lattices, and may contribute to the discovery of new strongly correlated states of matter.
\end{abstract}

\maketitle

\section{\label{Sec1}Introduction}

Moir\'{e} lattices arise when two similar periodic structures are stacked at a twist angle, resulting in long-period moir\'{e} patterns~\cite{RN1,RN2,RN3,RN4,RN5}. This overlay fundamentally alters the system's original quantum mechanical properties, offering an exciting platform that introduces an additional degree of freedom to manipulate quantum phases, thereby enhancing our understanding of correlated quantum matter~\cite{RN7,kennes2021moire,wang2019properties,lu2019superconductors,kennes2020one}.

In 2018, Cao et al. fabricated a graphene moir\'{e} lattices by stacking two monolayer graphene sheets with interlayer interactions and tuning the twist angle between them\cite{RN8,RN9}. They experimentally demonstrated that a flat band appears near the Fermi level at a special twist angle, referred to as the ``magic angle". This flat band leads to a vanishing Fermi velocity at the Dirac point and a high density of states, resulting in the emergence of correlated insulating states and unconventional superconductivity in twisted bilayer graphene (TBG). Since then, graphene-based moir\'{e} systems have emerged as a central focus of research, as they offer a clean and versatile platform for studying strongly correlated materials and provide valuable insights into the mechanisms of superconductivity~\cite{RN10,RN11,RN12,RN13,yankowitz2019tuning,balents2020superconductivity}. Recently, many exotic phases have been observed in TBG, including the quantum anomalous Hall effect  \cite{RN14,RN15}, the valley Hall effect~\cite{RN15,RN16} and fractional Chern insulators \cite{RN17,RN18,RN19}. Theoretical challenges in understanding these correlation-driven phenomena in TBG primarily arise from the presence of flat bands and their associated topological properties~\cite{RN20,RN21,RN22,wang2024proposal}.

Given the limitations on accessible physical parameters, such as the twist angle and interlayer coupling strength, in two-dimensional (2D) materials, extending twisted bilayer van der Waals structures to new platforms is highly desirable, as it offers promising opportunities to explore twistronics beyond TBG. In photonics, moir\'{e} lattices have been realized in photonic crystals and photonic lattices, which benefit from highly controllable lattice structures and symmetries~\cite{RN28,RN30,RN31}. These systems have enabled the observation of exotic phenomena such as light localization and high-quality nanocavities~\cite{RN6,RN23,RN24}. In ultracold atoms, a twisted bilayer optical lattice has been realized by our group~\cite{RN25}, offering significant advantages for studying moir\'{e} physics with unprecedented tunability \cite{mcdonald2019superresolution,gonzalez2019cold,DebrajRakshit2020,luo2021spin,ZheyuShi2024,gall2021competing,tarruell2012creating,soltan2011multi, windpassinger2013engineering,bloch2008many}.
	
The moir\'{e} flat bands in TBG are critically influenced by both the twist angle and the interlayer coupling strength (ICS). In TBG, the ICS can be finely tuned by applying out-of-plane pressure~\cite{RN26,RN27,yankowitz2018dynamic,yankowitz2019tuning}. However, the range of this adjustment is quite limited, which poses challenges in studying the continuous transition from weak ICS to strong ICS in moir\'{e} physics. Here, based on the highly controllable interlayer coupling of ultracold atoms in a 2D twisted bilayer honeycomb optical lattice \cite{luo2021spin}, we investigate the rich moir\'{e} flat bands across a broad range of twist angles. We identify an isolated topological flat band at the Dirac point energy for a specific value of weak ICS, referred to as the "critical coupling." The value of this critical coupling depends on the twist angle. When the ICS is slightly increased beyond the critical coupling, the topological flat band touches both the upper and lower bands at the high-symmetry point $\Gamma_s$. As the ICS enters the strong coupling regime, multiple trivial flat bands emerge around the ``Dirac point energy bands". Here, we define the ``Dirac point energy bands" as the energy bands that form the Dirac point without interlayer coupling. These trivial bands remain flat as the ICS is further increased. We also find that at the critical coupling, the band at the lowest energy becomes flat as well, but it remains a topologically trivial flat band, in contrast to the topologically nontrivial flat band formed at the Dirac point energy. As the ICS continues to increase, the number of trivial flat bands gradually increases from lower to higher energy bands. The topological properties of the flat bands are analyzed by calculating the Wilson loop spectrum.

The rest of the paper is organized as follows. In Sec.~\ref{Sec2}, we present the theoretical model for describing dilute ultracold atomic gases in twisted bilayer honeycomb optical lattices. Specifically, Sec.~\ref{Sec3A} discusses the emergence of topological flat bands in the weak interlayer coupling regime, including the critical coupling as a function of the twist angle. In Sec.~\ref{Sec3B}, we analyze the appearance of trivial flat bands in the strong interlayer coupling regime. Finally, we discuss the experimental relevance of our results and provide concluding remarks in Sec.~\ref{Sec4}.

\section{\label{Sec2}Theoretical Model}

The twisted bilayer honeycomb optical lattice can be realized using synthetic dimension techniques previously employed in our experiment~\cite{RN25}. Initially, atoms are confined to a quasi-two-dimensional pancake-shaped potential by a deep trap along the $z$-axis. Subsequently, atoms in two different spin states are loaded into two independent honeycomb optical lattices with a relative twist angle. These two lattices selectively address atoms in different spin states, thereby forming a synthetic dimension that represents the bilayer structure. Interlayer coupling is introduced via a microwave field that coherently couples the two spin states.

The Hamiltonian of the system can be written as
\begin{equation}
H = \begin{pmatrix}
-\frac{\hbar^2}{2m_a}\nabla^2 + V_1 & \Omega_R \\
\Omega_R & -\frac{\hbar^2}{2m_a}\nabla^2 + V_2
\end{pmatrix},
\end{equation}
where  $m_a$ is the atomic mass, $\hbar$ is the reduced Planck constant, and  $\Omega_R$ denotes the ICS. The honeycomb optical lattice potentials $V_1$ and $V_2$ are defined as
\begin{align}
V_1&= V_0 \left[ \cos\left(\frac{3}{2}k_0x_1 + \frac{\sqrt{3}}{2}k_0y_1\right) \right. \cr\cr
&\left. + \cos\left(-\frac{3}{2}k_0x_1 + \frac{\sqrt{3}}{2}k_0y_1\right) + \cos\left(-\sqrt{3}k_0y_1\right) \right]\!\!,\cr\cr
V_2&= V_0 \left[ \cos\left(\frac{3}{2}k_0x_2 + \frac{\sqrt{3}}{2}k_0y_2\right) \right. \cr\cr
&\left. + \cos\left(-\frac{3}{2}k_0x_2 + \frac{\sqrt{3}}{2}k_0y_2\right) + \cos\left(-\sqrt{3}k_0y_2\right) \right]\!\!,
\end{align}
where the rotated coordinates are defined as $x_1=x\cos(\theta/2)+y\sin(\theta/2)$, $y_1=-x\sin(\theta/2)+y\cos(\theta/2)$, $x_2=x\cos(\theta/2)-y\sin(\theta/2)$ and $y_2=x\sin(\theta/2)+y\cos(\theta/2)$ with $x$ and $y$ being the spatial coordinates and $\theta$ the twist angle. The wave number of the lattice lasers is $k_0 = 2\pi/\lambda$, where $\lambda$ is the laser wavelength. The lattice depth $V_0$ is expressed in units of the recoil energy, defined as $E_r = \hbar^2k_0^2/(2m_a)$. The moir\'{e} lattice structure formed by $V_1$ and $V_2$ is illustrated in Figs. 1(a) and 1(b).

\begin{figure*}[t]
\centering
\includegraphics[width=1\linewidth]{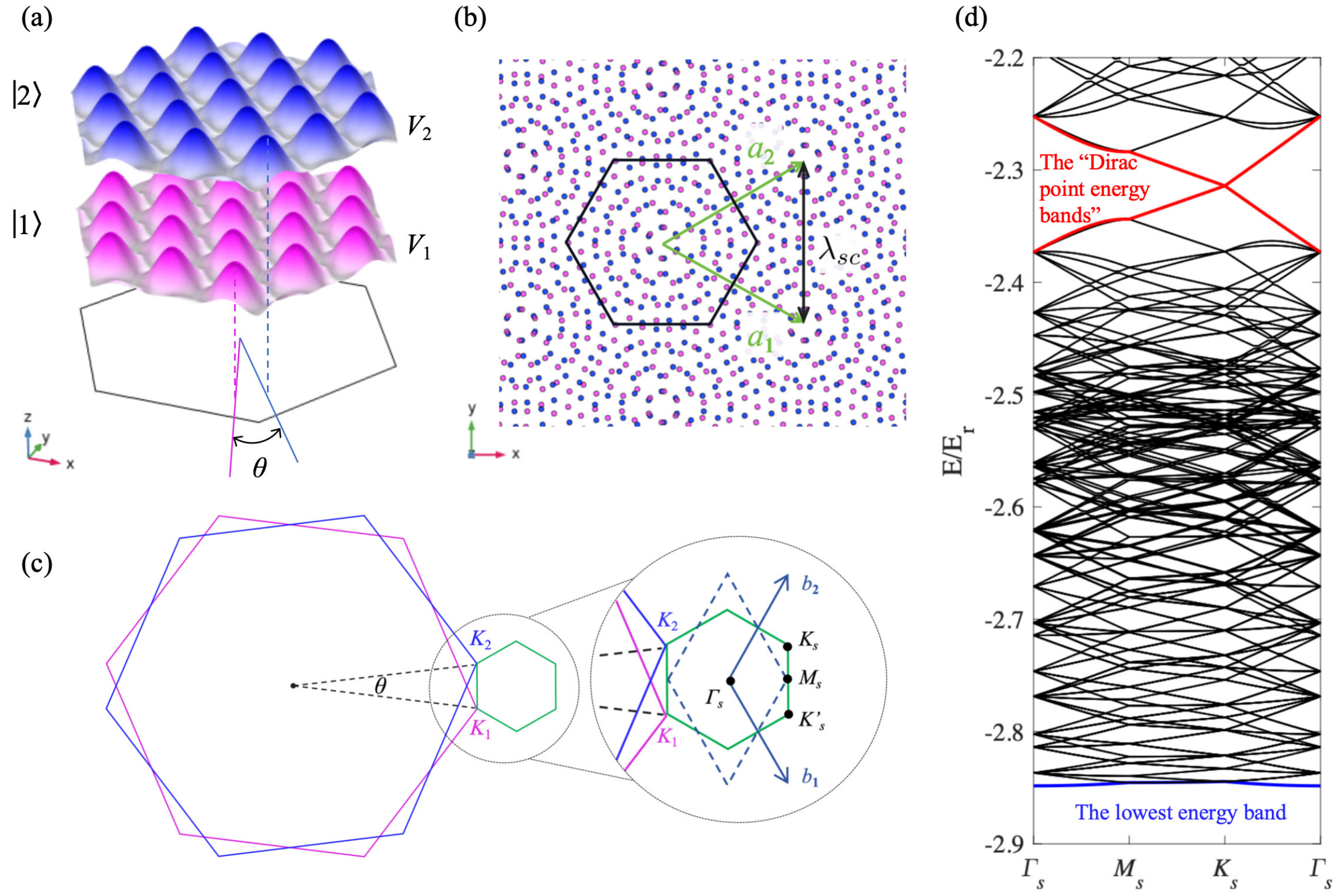}
\caption{(a) Moir\'{e} supercell formed by two sets of honeycomb optical lattices $V_1$ (magenta) and $V_2$ (blue) with a small twist angle $\theta$. The lattices $V_1$ and $V_2$ independently confine ultracold atoms in spin states $\left|1\right\rangle$ and $\left|2\right\rangle$, respectively. (b) Periodic structure of the twisted bilayer honeycomb optical lattice. The magenta and blue dots represent the minima of the optical potentials $V_1$ and $V_2$, respectively. $\mathbf{a_1}$ and $\mathbf{a_2}$ denote the lattice vectors, and $\lambda_{sc}$ is the period of the moir\'{e} supercell. (c) The first Brillouin zones of the monolayer honeycomb optical lattices (magenta and blue hexagons), along with the resulting moir\'{e} Brillouin zone (green hexagon). The high-symmetry points $K_1$, $K_2$, $\Gamma_s$, $K_s$, $M_s$ and $K_s'$ are labeled, and the reciprocal lattice vectors are denoted by $\mathbf{b_1}$ and $\mathbf{b_2}$. (d) Energy bands of the moir\'{e} lattice for $m=22$, $n=1$ ($\theta \approx 4.4085^\circ$), with $V_0 = 4E_r$ and $\Omega_R = 0$. The red lines highlight the ``Dirac-point energy bands", while the blue line indicates the lowest energy band.}
\label{fig:fig1}
\end{figure*}

A twisted bilayer honeycomb optical lattice can exhibit either a periodic (commensurate) or an aperiodic (incommensurate) structure, depending on the twist angle. The commensurate twist angles are determined as~\cite{AHCastro2007}
\begin{equation}
\theta = \arctan\left(\frac{\sqrt{3} b}{2a + b}\right),
\end{equation}
where the integers $a$ and $b$ are defined based on the values of $m$ and $n$, which are coprime natural numbers satisfying $m > n$. If $(m - n)/3$ is an integer, then $a = (m^2 - n^2)/3$ and $b = (2mn + n^2)/3$; otherwise, $a = m^2 - n^2$ and $b = 2mn + n^2$. Accordingly, the period of the moir\'{e} supercell, denoted by $\lambda_{sc}$  is given by $\lambda_{sc}=\sqrt{(m^2+n^2+mn)/3}=n\lambda_{mo}$ when $(m-n)/3$ is an integer, and by $\lambda_{sc}=\sqrt{m^2+n^2+mn}=\sqrt{3}n\lambda_{mo}$ when $(m-n)/3$ is a non-integer. Here, $\lambda_{mo}=\lambda/[3\sin(\theta/2)]$ represents the period of the moir\'{e} pattern.

The first Brillouin zones of the two honeycomb optical lattices, along with the moir\'{e} Brillouin zone defined by the new moir\'{e} periodicity, are illustrated in Fig. 1(c). For small twist angles, the moir\'{e} Brillouin zone becomes significantly smaller than that of a monolayer honeycomb optical lattice. As a result, each energy band of the monolayer lattice is folded into multiple bands within the moir\'{e} Brillouin zone. The number of folded bands is given by $c=(m^{2}+n^{2}+mn)/3$ when $(m-n)/3$ is an integer, and by $c= m^{2}+n^{2}+mn$ otherwise. For instance, the original $s$ band of the monolayer lattice is folded into the first $c$ bands of the moir\'{e} lattice, and the original $p$ band is folded into the moir\'{e} bands ranging from $c+1$ to $2c$.

For simplicity, we focus on the case where \(n = 1\) and \((m - n)/3\) is an integer, which corresponds to a supercell containing a single period of the moir\'{e} pattern, as illustrated in Fig.~1(b). In this scenario, the energy bands exhibit the lowest possible degeneracy. The Dirac point appears within the folded bands, ranging from $2c-1$ to $2c+2$ and we refer to these four bands as the ``Dirac point energy bands". Specifically, for $m=22$, $n=1$ (corresponding to $\theta\approx4.4085^\circ$) and $\Omega_R=0$, the band structure of the twisted bilayer honeycomb optical lattice is shown in Fig.~1(d). As indicated by the red lines, the Dirac point energy bands are located between the $337$-th and $340$-th bands of the moir\'{e} lattice. All numerical calculations in this work were performed using the finite element method implemented in COMSOL Multiphysics.

\begin{figure*}[t]
\centering
\includegraphics[width=1\linewidth]{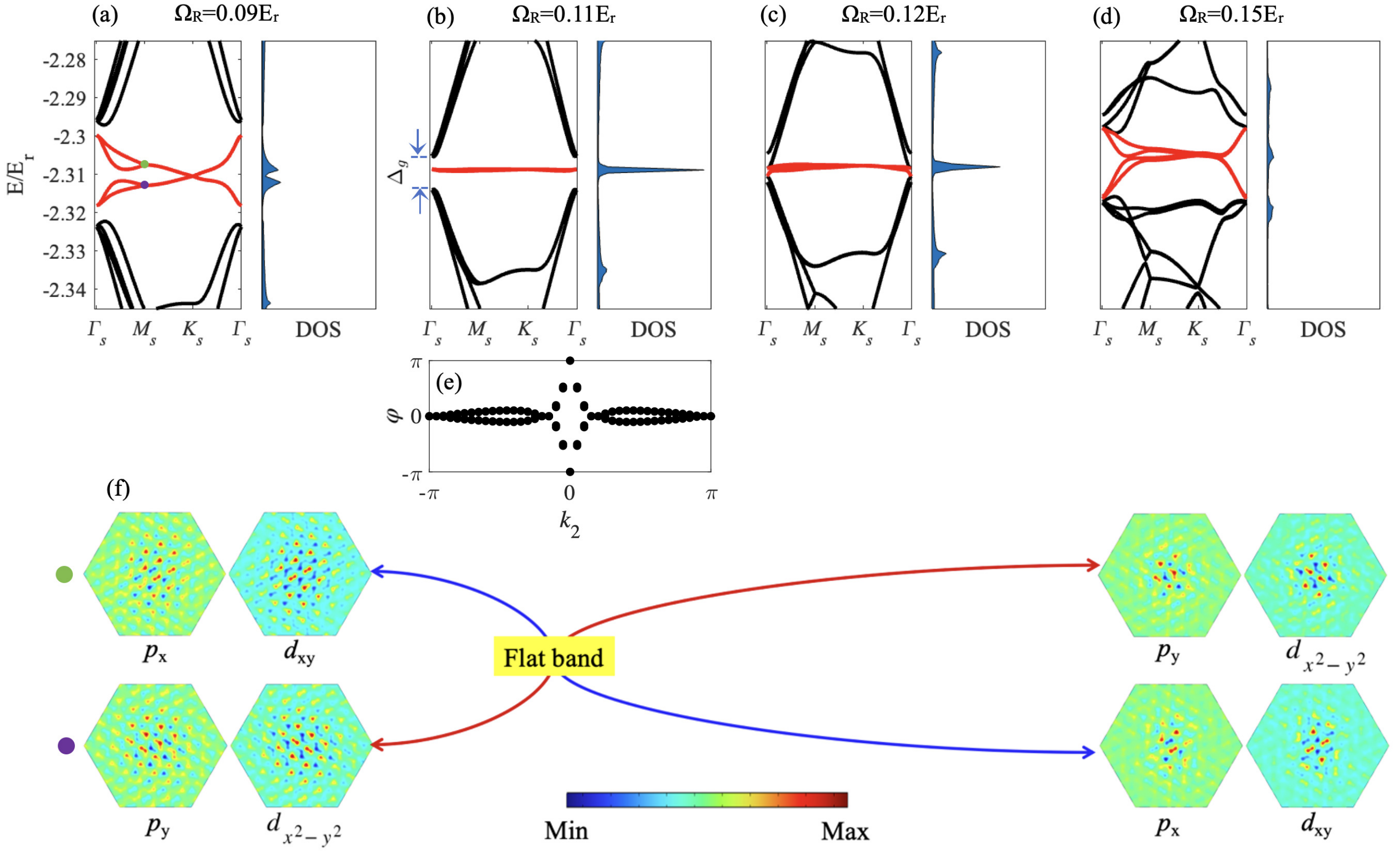}
\caption{(a)-(d) Energy bands and corresponding density of states near the Dirac points of the twisted bilayer honeycomb optical lattice for increasing interlayer coupling strength $\Omega_R=0.09E_r, 0.11E_r, 0.12E_r$ and $0.15E_r$, respectively. (e) Wilson loop of the topological flat band shown in panel (b). (f) Real part of the eigenstates at the $M_s$ point of the moir\'{e} Brillouin zone for interlayer coupling strength $\Omega_R = 0.09E_r$ (left panel) and $0.15E_r$ (right panel). The twist angle is fixed at $\theta \approx 4.4085^\circ$, and the lattice depth is set to $V_0 = 4E_r$.}
\label{fig:fig2}
\end{figure*}

\section{\label{Sec3}Flat bands induced by interlayer coupling}

The interlayer coupling between atoms in the two twisted optical honeycomb lattices plays a crucial role in determining the energy band structure. We investigate the energy bands in both the weak and strong interlayer coupling regimes, and discuss the emergence of flat bands as well as their topological properties.

\subsection{\label{Sec3A}The weak interlayer coupling regime}

We compute the energy band structure for weak ICS near the Dirac point, as shown in Fig. 2. For a fixed commensurable twist angle of $\theta=\arctan\frac{15\sqrt{3}}{337}\approx4.4085^\circ$, the energy bands and corresponding density of states for different ICS values are presented in Figs. 2(a)-2(d). Compared with the zero ICS case shown in Fig. 1(d), one can clearly see from Fig. 2(a) that the interlayer coupling significantly reduces the energy width of the Dirac cone and opens two band gaps between the ``Dirac point energy bands" and their adjacent upper and lower bands. The continued existence of the Dirac point, along with the emergence of band gaps between the ``Dirac point energy bands" and their adjacent bands, can be understood as follows: The Dirac point degeneracy originates from the sublattice symmetry between the A and B sites in a single-layer honeycomb lattice. This type of degeneracy is protected by both time-reversal symmetry and spatial inversion symmetry (sublattice exchange). In the regime of weak interlayer coupling, the Dirac points in the bilayer system remain largely governed by the intrinsic sublattice symmetry of each individual layer, thus the interlayer coupling is insufficient to break this symmetry-protected degeneracy. In contrast, the degenerate crossings between the ``Dirac point energy bands" and their adjacent bands arise from the band folding in momentum space of single-layer graphene energy bands into the mini Brillouin zone of the moir\'{e} superlattice. These degeneracies are purely geometric and accidental, not protected by symmetry. Such accidental degeneracies are generally lifted by even weak interlayer coupling.

As the ICS approaches a critical value, the group velocity at the Dirac point nearly vanishes, the hybridized energy bands become flat, and a pronounced peak emerges in the density of states, as shown in Fig. 2(b). This specific value of ICS, which minimizes the bandwidth, is referred to as the ``critical coupling". When the ICS is slightly increased beyond the critical coupling, the flat band touches the upper and lower bands at the high-symmetry $\Gamma_s$ point, as shown in Fig. 2(c). In this regime, the previously isolated flat band evolves into a singular flat band with band touching. Upon further increasing the ICS, the bandwidth broadens and the flatness of the band is lost, as illustrated in Fig. 2(d).

To gain deeper insight into the formation of the flat band, we analyze the energies and Bloch modes of the ``Dirac point energy bands" at the $M_s$ point of the moir\'{e} Brillouin zone, which is associated with Van Hove singularities~\cite{RN8}. As shown in Figs.~2(a)-2(d), interlayer coupling induces energy splitting at the $M_s$ point, bringing the energies of the four Bloch modes from the ``Dirac point energy bands" closer to that of the Dirac point. As the ICS increases to the critical coupling, these four Bloch modes become degenerate with the Dirac point energy, resulting in the formation of flat bands. When the ICS increases further, the energies of the four Bloch modes split again, lifting the degeneracy and resulting in the disappearance of the flat band. Meanwhile, we observe an orbital exchange among the Bloch modes of the ``Dirac point energy bands" before and after the band flattens, as illustrated in Fig. 2(f). Before the band becomes flat, the upper two bands host Bloch modes with $p_x$-like and $d_{xy}$-like orbitals, while the lower two bands contain $p_y$-like and $d_{x^2-y^2}$-like orbitals. After the flat band disappears due to increased ICS, this orbital distribution is reversed: the upper two bands exhibit $p_y$-like and $d_{x^2-y^2}$-like orbitals, while the lower two bands exhibit $p_x$-like and $d_{xy}$-like orbitals.

The topological properties of the flat bands can be characterized by examining the Wilson loop spectrum~\cite{RN21}. According to the formalism developed by Wilczek and Zee~\cite{111}, the time evolution of quantum states in an adiabatic system is described by the following path-ordered integral
\begin{equation}
\hat{W}_{\mathbf{k}(0) \rightarrow \mathbf{k}(t)} = \mathcal{P} \exp \left[ i \oint_{C} \hat{A}(\mathbf{k}) \cdot d\mathbf{k} \right],
\end{equation}
where $\mathcal{P}$ denotes path ordering, and the integration path $C$ extends from $\mathbf{k}(0)$ to $\mathbf{k}(t)$ in reciprocal space. Here, $\hat{A}(\mathbf{k})$ is the non-Abelian Berry connection, which captures the local geometric properties of the state space. In a periodic system, the Bloch state for the $n$-th band at a wave vector $\mathbf{k}$ is given by $|\psi_{n,\mathbf{k}}(\mathbf{r})\rangle = e^{i\mathbf{k} \cdot\mathbf{r}}\left|{u_{n,\mathbf{k}}}\right\rangle$. The matrix elements of the non-Abelian Berry connection are determined by the periodic part of the wave function $\left|{u_{n,\mathbf{k}}}\right\rangle$, and are calculated as $\left[ A(\mathbf{k}) \right]_{nn'} = i \langle u_{n,\mathbf{k}} \left| \nabla_{\mathbf{k}} \right| u_{n',\mathbf{k}} \rangle$.

We parametrize the wave vector as $\mathbf{k}=\frac{k_1}{2\pi}\mathbf{b_1}+\frac{k_2}{2\pi}\mathbf{b_2}$ in the rhombus-shaped moir\'{e} Brillouin zone, where $\mathbf{b_1}$ and $\mathbf{b_2}$ are the reciprocal lattice vectors, as shown in Fig. 1(c). For calculating the Wilson loop, we choose a path with a fixed $k_2$ in the moir\'{e} Brillouin zone, which is parallel to the $\mathbf{b_1}$ direction. In the case of an isolated energy band, the Wilson loop can be computed as
\begin{equation}
W_{k_2} = \prod_{i=1}^{j} \left\langle u_{n,k_1^{i},k_2} \middle| u_{n,k_1^{i+1},k_2} \right\rangle,
\end{equation}
where $j$ is sufficiently large to describe the infinite lattice limit ${j \to \infty}$. The Berry phase for characterizing the topological property of a single band is then given by
\begin{equation}
\varphi_{k_2}= \text{Im} [\ln(W_{k_2})],
\end{equation}
For $N$ degenerate bands, the Wilson loop matrix can be calculated as
\begin{equation}
\hat{W}_{k_2} = \prod_{i=1}^{j} \hat{M}^{(k_1^{i},k_2), (k_1^{i+1},k_2)},
\end{equation}
where $\hat{M}$ is a $N\times N$ matrix, and its matrix elements are given by
\begin{equation}
M_{nn'}^{(k_1^i,k_2),(k_1^{i+1},k_2)} = \left\langle u_{n,k_1^i,k_2} \middle| u_{n',k_1^{i+1},k_2} \right\rangle,
\end{equation}
with $n, n' \in \{1, \ldots, N\}$. Then the Berry phases, which characterize the topological properties of multiple bands, are obtained from the eigenvalues of the Wilson loop matrix as
\begin{equation}
\varphi_{k_2}^{n} = \text{Im}[\ln(w_{k_2}^{n})],
\end{equation}
where $w_{k_2}^{n}$ is the $n$-th eigenvalue of the matrix $\hat{W}_{k_2}$.

It is worth noting that the Berry phase is related to the Wannier center $c$ \cite{RN32}, which represents the centroid of the maximally localized Wannier function~\cite{marzari1997maximally}. When $c = \pm 0.5$, the centroid of the maximally localized Wannier function lies at the edge of the primitive cell, whereas for $c = 0$, it is located at the center of the primitive cell. In general, the Wannier function is obtained by performing a Fourier transform of the Bloch function over the two-dimensional Brillouin zone. Here, we get the Wannier function by applying a Fourier transform only along the $k_1$ direction for a fixed $k_2$. As a result, the Wannier center $c$ becomes a function of $k_2$, and is related to the Berry phase by $c(k_2)=\varphi(k_2)/2\pi$. The evolution of the Wannier center with respect to $k_2$ directly reflects the topological properties of the band. For a topologically nontrivial band, the Wannier center is delocalized, and the Berry phase spans the full range of $[-\pi, \pi]$. In contrast, for a topologically trivial band, the Wannier center is localized, corresponding to a localized spectrum of the Berry phase.

For the flat band shown in Fig.~2(b), the Wilson loop is calculated and presented in Fig.~2(e). It is evident that the Wilson loop, characterized by a winding number $w = 1$, spans the entire moir\'{e} Brillouin zone, indicating the nontrivial topology of the flat band. Previous studies have shown that topological flat bands originating from ``Dirac point energy bands" in twisted bilayer systems exhibit a distinct type of topology known as fragile topology~\cite{RN21,po2019faithful}. Fragile topological phases typically do not feature gapless edge states, thereby violating the conventional bulk-boundary correspondence, and hold great potential for modeling and engineering unconventional correlated materials~\cite{song2020twisted,peri2020experimental}.

To illustrate the influence of the twist angle and ICS on the bandwidth in detail, we calculate the bandwidth $\Delta E$ as a function of $\Omega_R$ for various twist angles $\theta$. As shown in Fig.~3(a), for a fixed twist angle, the bandwidth significantly decreases with increasing ICS, reaching a minimum at a critical coupling strength. This indicates that even for relatively large twist angles (e.g., $\theta \approx 5.0858^\circ$ with $m = 19$), the topological band near the Dirac point can still be effectively flattened. This provides a more flexible approach to realizing topological flat bands, in contrast to the narrow magic angle range ($\sim 1^\circ$) required in twisted bilayer graphene. Moreover, Fig.~3(a) shows that as the twist angle increases, a larger critical coupling strength is required to flatten the topological band. When the twist angle $\theta\gtrsim7.341^\circ$ ($m\leq13$), the critical coupling is no longer sufficient to induce flat bands, as demonstrated in Fig.~3(b). We further compute the relative bandwidth $\Delta E / \Delta_g$ at the critical coupling for different twist angles, where $\Delta_g / 2$ denotes the average band gap between the flat band and its neighboring upper or lower band, as shown in Fig.~3(c). It is evident that for $\theta \lesssim 6.009^\circ$ ($m \geq 16$), the relative bandwidth remains small ($\Delta E / \Delta_g < 0.2$), whereas for $\theta \gtrsim 7.341^\circ$ ($m \leq 13$), the relative bandwidth becomes large ($\Delta E / \Delta_g > 0.5$). In addition, we investigate the dependence of the critical coupling strength on the lattice depth $V_0$, and find that increasing $V_0$ reduces the critical coupling strength required for the emergence of topological flat bands. The corresponding critical coupling strengths for various twist angles $\theta$ and lattice depths $V_0$ are summarized in Table~I.

\begin{figure}[t]
\centering
\includegraphics[width=1\linewidth]{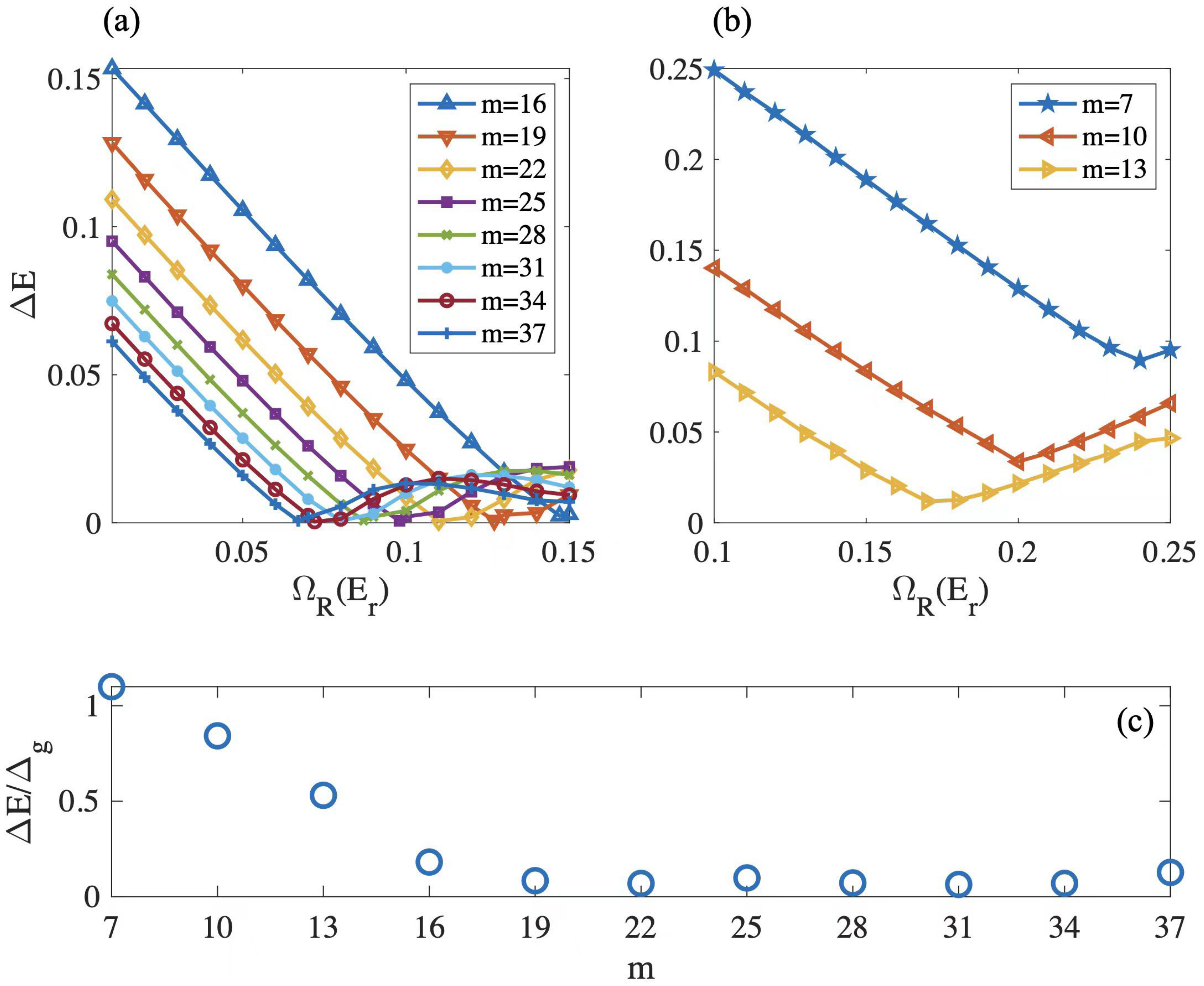}
\caption{The bandwidth of the ``Dirac point energy bands" as a function of the interlayer coupling strength for various twist angles. (a) The cases with $n=1$, $m=16, 19, 22, 25, 28, 31, 34$, and 37, corresponding to twist angles $\theta\approx6.009^{\circ}, 5.0858^{\circ}, 4.4085^{\circ}, 3.8902^{\circ}, 3.481^{\circ}, 3.1497^\circ, 2.8759^{\circ}$, and $2.6459^{\circ}$. (b) The cases with $n=1, m=7$, 10, and 13, corresponding to $\theta\approx13.174^{\circ}, 9.43^{\circ}$, and $7.341^{\circ}$. (c) The relative bandwidth $\Delta E/\Delta_g$ at the critical coupling as a function of twist angle. The lattice depth is fixed at $V_0 = 4E_r$ throughout.}
\label{fig:fig3}
\end{figure}

\begin{table}[t]
\begin{center}
\caption{Critical coupling strengths for different twist angles. The twist angles $\theta$ are derived from commensurate configurations with $n = 1$ and $(m - n)/3$ being an integer. $\Omega_R^{a} (E_r)$ and $\Omega_R^{b} (E_r)$ denote the critical coupling strengths for lattice depths $V_0 = 4E_r$ and $6E_r$, respectively.}
\begin{ruledtabular}
\begin{tabular}{cccc}
\textit{m} & \textrm{$\theta(^{\circ})$} & \textrm{$\Omega_R^{a}(E_r)$} & \textrm{$\Omega_R^{b}(E_r)$} \\
\hline
19 & 5.0858 & 0.127 & 0.115\\
22 & 4.4085 & 0.110 & 0.101\\
25 & 3.8902 & 0.098 & 0.089\\
28 & 3.4810 & 0.087 & 0.080\\
31 & 3.1497 & 0.079 & 0.072\\
34 & 2.8759 & 0.072  & 0.066\\
37 & 2.6459 & 0.067& 0.061\\
40 & 2.4500 & 0.062 & 0.056\\
43 & 2.2811 & 0.058 & 0.052\\
46 & 2.1339 & 0.054 & 0.049\\
49 & 2.0046 & 0.050 & 0.046\\
\end{tabular}
\end{ruledtabular}
\end{center}
\end{table}

\subsection{\label{Sec3B}The strong interlayer coupling regime}

As discussed above, when the ICS increases further, the topological flat bands around the Dirac point gradually lose their flatness. In contrast, the lowest energy bands behave quite differently. At the critical coupling, the lowest band becomes an isolated and nearly flat single band, as exemplified by $\theta \approx 4.4085^\circ$ ($m = 22$), as shown in Fig.~4(a). As the ICS continues to increase, this lowest energy band becomes increasingly flat, as illustrated in Fig.~4(b). Furthermore, an increasing number of flat bands emerge sequentially from lower to higher energies and remain flat with increasing ICS, as shown in Figs.~4(b) and 4(c). The detailed evolution of the bandwidth $\Delta E$ for the lowest ten bands as a function of ICS is shown in Fig.~4(d).

The emergence of flat bands results in real-space localization of the corresponding eigenstates. The real parts of the eigenstates associated with the lowest ten energy bands are displayed in the insets of Fig.~4(c). Specifically, the 1st, the degenerate 2nd and 3rd, the 4th, the degenerate 5th and 6th, the degenerate 7th and 8th, the 9th, and the 10th bands are composed of localized orbitals resembling $1s$, $2p_x$ ($2p_y$), $2s$, $3p_x$ ($3p_y$), $3d_{xy}$ ($3d_{x^2 - y^2}$), $4f_{x(x^2 - 3y^2)}$, and $4f_{y(3x^2 - y^2)}$ orbitals, respectively. It is worth noting that although the 9th and 10th bands are nearly degenerate, they can still be regarded as two isolated bands. For a given flat band, we find that the periodic part of the Bloch eigenstates is independent of the wave vector. This implies a vanishing quantum metric and Berry curvature, indicating a topologically trivial band structure. By calculating the Wilson loop, we have verified that all of the lowest ten flat bands are topologically trivial, as shown in Fig.~4(e).

\begin{figure*}
\centering
\includegraphics[width=1\linewidth]{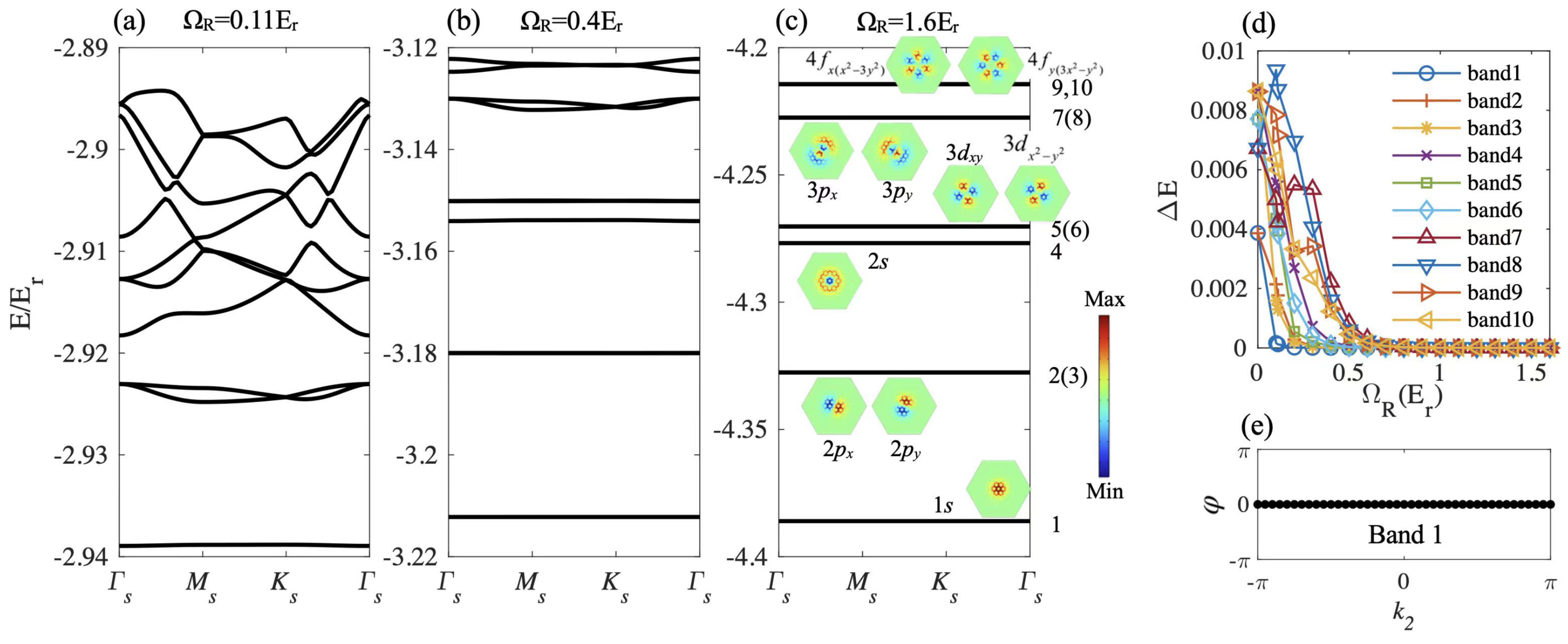}
\caption{Trivial flat bands emerging within the lowest ten energy bands in the strong interlayer coupling regime. (a)-(c) The lowest ten energy bands for $n = 1$, $m = 22$ (corresponding to $\theta \approx 4.4085^\circ$), with $\Omega_R = 0.11E_r$, $0.4E_r$, and $1.6E_r$, respectively. Insets in (c) show the real parts of the eigenstates for the lowest ten bands. Specifically, the 1st, degenerate 2nd (3rd), 4th, degenerate 5th (6th), degenerate 7th (8th), 9th, and 10th bands correspond to localized orbitals resembling $1s$, $2p_x$/$2p_y$, $2s$, $3p_x$/$3p_y$, $3d_{xy}$/$3d_{x^{2}-y^{2}}$, $4f_{x(x^{2}-3y^{2})}$, and $4f_{y(3x^{2}-y^{2})}$, respectively. (d) Evolution of the bandwidth $\Delta E$ of the lowest ten bands as a function of the interlayer coupling strength $\Omega_R$. (e) Wilson loop of the lowest band with trivial topology.}
\label{fig:fig4}
\end{figure*}

Further increasing the ICS into the strong coupling regime leads to the emergence of additional flat bands at higher energies. Notably, for sufficiently strong ICS, the ``Dirac point energy bands" become flat once again. In Figs.~5(a)-(c), we display the 337th to 340th bands along with their nearby bands at $\Omega_R = 2E_r$, $6E_r$, and $10E_r$, respectively. The 337th to 340th energy bands, highlighted in red, correspond to the ``Dirac point energy bands" at $m = 22$. It is evident that flat bands reappear near the Dirac point when $\Omega_R$ exceeds $6E_r$. In contrast to the topologically nontrivial flat bands that emerge near the Dirac point at the critical coupling, the flat bands formed in the strong interlayer coupling regime are all topologically trivial. This is confirmed by calculating the Wilson loops, as shown in Figs.~5(d)-(f). As the ICS approaches infinity, the system effectively reduces to a single layer (or single component) subjected to a twisted optical lattice~\cite{RN25}, which has been studied experimentally in photonic systems~\cite{RN6,RN24,RN29,huang2016localization,luan2023reconfigurable}.

\begin{figure}
\centering
\includegraphics[width=1\linewidth]{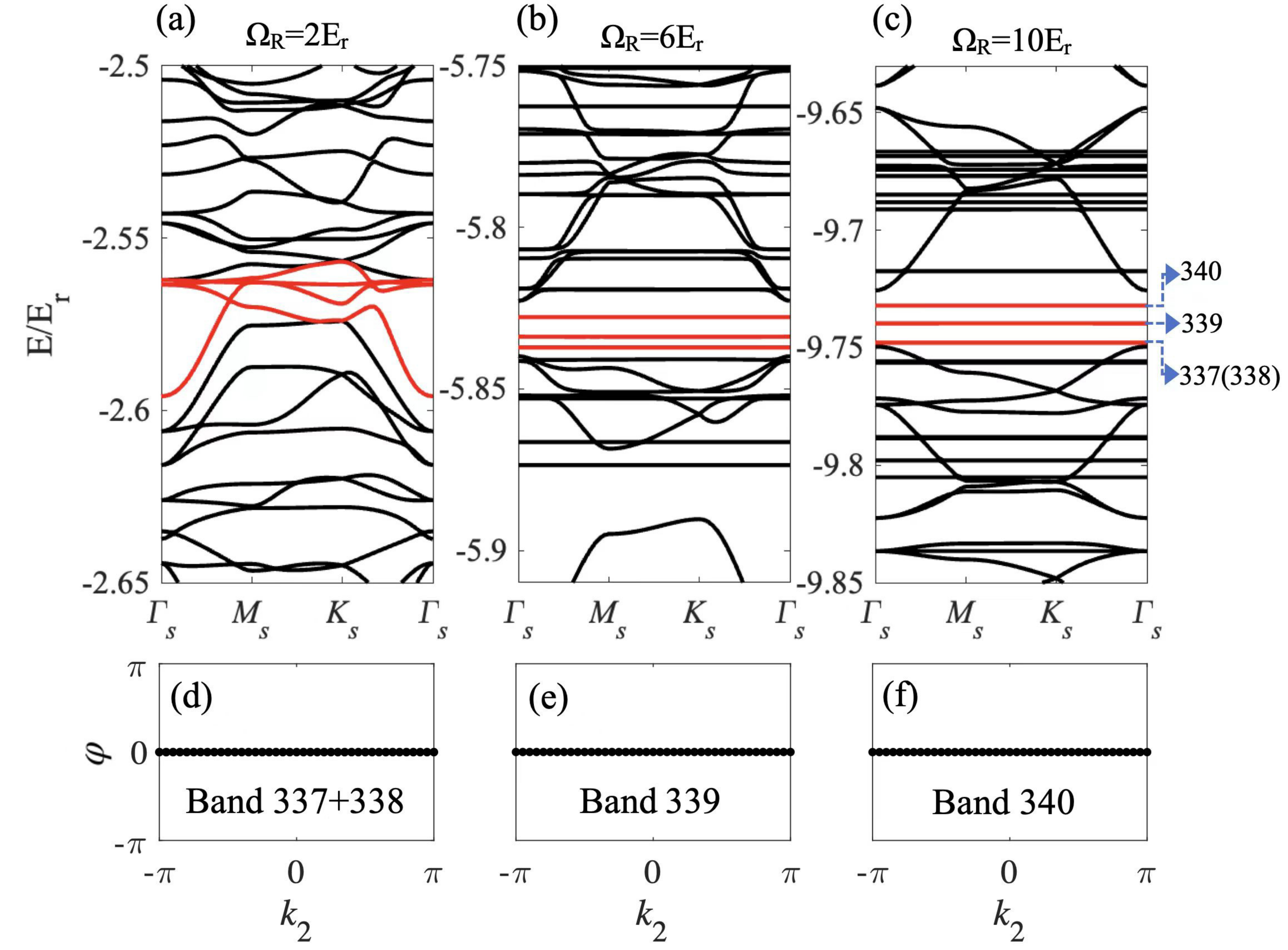}
\caption{Trivial flat bands formed by the ``Dirac point energy bands" in the strong interlayer coupling regime. (a)-(c) The ``Dirac point energy bands" (337th to 340th bands for m = 22) highlighted in red, along with nearby energy bands, for $\Omega_R=2E_r, 6E_r$ and $10E_r$, respectively. (d)-(f) Wilson loops of the trivial flat bands.}
\label{fig:fig5}
\end{figure}

\section{\label{Sec4}Discussion and Conclusion}

In realistic experiments, the topological and trivial flat bands can be detected by employing amplitude modulation spectroscopy~\cite{TEsslinger2004,KSengstock2018} and spin-injection radio-frequency spectroscopy~\cite{DSJin2008,JZhang2012,Zwierlein2012}, which can induce inter-band transitions and allow for the reconstruction of the band dispersion. To observe the topological properties of flat bands near the Dirac point, atoms can be transported through reciprocal space by uniformly accelerating the lattice via a linear frequency sweep of the lattice beams~\cite{DStamper-Kurn2022,JZhang2024}. This generates a constant inertial force, enabling access to the Wilson line regime, where the dynamics are governed entirely by geometric effects. The Wilson line can then be measured by tracking changes in the band populations~\cite{USchneider2016}.

One can use the clock states $^{1}S_0$ and $^{3}P_0$ of alkaline-earth (-like) atoms as the two pseudospin states for the twisted bilayer optical lattices~\cite{luo2021spin,JYe2017}. The $J=0\rightarrow J=0$ forbidden transitions result in an extremely narrow natural linewidth (approximately 1 mHz), and the transition is insensitive to magnetic field perturbations, making it beneficial for the high-precision detection of the energy bands. Considering the ultra-narrow gap between the topological flat band and neighboring bands, which may exceed the atomic thermal energy at finite temperature, experimental detection becomes challenging. A promising solution is to enhance the gap by introducing spatially dependent interlayer coupling~\cite{RN5,ZYMneg2024}.

While the topological flat band at the single-particle level is similar to that in twisted bilayer graphene, the strongly correlated effects induced by the flat band in ultracold atoms will be fundamentally different from those in graphene. This is attributed to the unique and highly tailorable many-body interactions in ultracold atomic systems~\cite{CChin2010,TPfau2023,IBloch2012}. For example, the attractive s-wave interaction between the Fermi atoms from different layers may induce a Larkin-Ovchinnikov superfluid at the topological flat bands~\cite{luo2021spin}. This phase exhibits a nonzero pairing momentum and a staggered superfluid density distribution in real space, and can be experimentally observed through time-of-flight measurements and in-situ imaging. In addition, exotic Mott insulating phases or Bose glass phases may also emerge due to the interplay between the flat bands and the interlayer atomic interactions~\cite{BDeMarco2016}.

In summary, we have studied the energy bands of twisted bilayer honeycomb optical lattices for ultracold atoms, with highly controllable interlayer coupling. We find that topologically nontrivial flat bands can form at a critical coupling in the weak interlayer coupling regime, whereas topologically trivial flat bands emerge in the strong interlayer coupling regime. With increasing interlayer coupling strength, the topologically nontrivial flat bands lose their flatness, whereas the topologically trivial flat bands become progressively flatter. For large twist angles, topological flat bands around the Dirac point can also be formed by tuning the interlayer coupling strength.
This goes beyond the limitations of the twist angle in twisted bilayer graphene. Our research reveals the physical principles of moir\'{e} bands more deeply and provides a universal guideline for applying moir\'{e} bands in other fields, including optics, acoustics, and condensed matter.

\section*{ACKNOWLEDGMENTS}

This research is supported by National Key Research and Development Program of China (Grants No. 2022YFA1404101, and No. 2021YFA1401700), Innovation Program for Quantum Science and Technology (Grants  No. 2021ZD0302003), National Natural Science Foundation of China (Grants No. 12488301, No. 12034011, No. U23A6004, No. 12474266, No. 12474252, No. 12374245, No. 12322409, and No. 92065108).

\end{document}